# Cytomegalovirus Antigenic Mimicry of Human Alloreactive Peptides: A Potential Trigger for Graft versus Host Disease

Charles Hall [1], Vishal Koparde [2], Max Jameson-Lee [1], Abdelrhman Elnasseh [1], Allison Scalora [1], Jared Kobulnicky [1], Myrna Serrano [2], Catherine Roberts [1], Gregory Buck [2,3], Micheal Neale [4], Daniel Nixon [5], and Amir Toor [1].

[1] Bone Marrow Transplant Program, Massey Cancer Center, Virginia Commonwealth University, Richmond, VA;

[2] Center for the Study of Biological Complexity, Virginia Commonwealth University, Richmond, VA;

[3] Department of Microbiology and Immunology, Virginia Commonwealth University, Richmond, VA;

[4] Departments of Psychiatry and Human & Molecular Genetics, Virginia Commonwealth University, Richmond, VA;

[5] Division of Infectious Diseases, Virginia Commonwealth University, Richmond, VA.

Address correspondence to

Amir A. Toor, MD
Professor of Medicine
VCU Massey Cancer Center
Bone Marrow Transplant Program
1300 East Marshall Street
PO Box 980157
Richmond, VA 23298-0157
Phone: 804-828-4360
Email: amir.toor@vcuhealth.org

**Abstract**

The association between human cytomegalovirus (hCMV) reactivation and the development of graft-versus-host-disease (GVHD) has been observed in stem cell transplantation (SCT). Seventy seven SCT donor-recipient pairs (DRP) (HLA matched unrelated donor (MUD), n=50; matched related donor (MRD), n=27) underwent whole exome sequencing to identify single nucleotide polymorphisms (SNPs) generating alloreactive peptide libraries for each DRP (9-mer peptide-HLA complexes); Human CMV CROSS (Cross-Reactive Open Source Sequence) Database was compiled from NCBI; HLA class I binding affinity for each DRPs HLA was calculated by NetMHCpan 2.8 and hCMV- derived 9-mers algorithmically compared to the alloreactive peptide-HLA complex libraries. Short consecutive (≥6) amino acid (AA) sequence homology matching hCMV to recipient peptides was considered for HLA-bound-peptide (IC50<500nM) cross reactivity. Of the 70,686 hCMV 9-mers contained within the hCMV CROSS database, 29,658.8 ± 9038.5 were found to match MRD DRP alloreactive peptides and 52,910.2 ± 16121.8 matched MUD DRP peptides (Student's T-test, $p<0.001$). *In silico* analysis revealed multiple high affinity, immunogenic CMV-Human peptide matches (IC50<500 nM) expressed in GVHD-affected tissue-specific manner (proteins expressed at ≥10 RPKM). hCMV+GVHD was found in 18 patients, 13 developing hCMV viremia before GVHD onset with a subset analysis of 7 instances of hCMV viremia prior to acute GVHD onset (n=3), chronic GVHD (n=2) and acute + chronic GVHD (n=2) indicating cross reactive peptide expression within affected organs . We propose that based on our analysis and preliminary clinical correlations that hCMV immune cross-reactivity may cause antigenic mimicry of human alloreactive peptides triggering GVHD.

**Introduction**

Human cytomegalovirus and other viral infections pose a significant hurdle to successful stem cell transplantation, affecting morbidity and mortality rates among the immunocompromised populations the world over. [1,2] Human CMV seropositivity has been estimated in 50-80% of the population in the United States by age 40. [3] While the majority of infected patients are asymptomatic due to viral latency after initial viral clearance, SCT patients with newly reconstituted immune systems exhibit rates of reactivation approximately 30-65% in multiple studies. [4-7] Furthermore, hCMV reactivation tends to be associated with the incidence of another serious complication of SCT, i.e., GVHD. Although many groups have historically separated GVHD incidence from hCMV infection data, even by complete exclusion in clinical trial design, there has been growing evidence that these diseases are not mutually exclusive and are likely linked in a number of ways beyond treatment. [8,9]

GVHD pathophysiology is driven by donor T cell mediated alloreactivity, directed against recipient mHA. These are recipient-derived oligopeptides presented on HLA molecules, and result from coding nucleotide sequence variation. Whole exome sequencing has been used to identify the entire library of nucleotide variation that exists between HLA matched SCT donors and recipients. [10] From these data it is possible to derive, *in silico*, all the *potential* alloreactive peptides which bind the relevant HLA (class I for example) in a donor-recipient-pair (DRP), and in turn this metric may be used to estimate a patient's given potential to develop alloreactivity, or GVHD. [11] Aside from helping understand GVHD pathophysiology, this library of recipient-derived mHA-HLA complexes may also be used to interrogate the relationship between pathogen antigens, as well as tumor antigens, and the disease states that may result, in particular the development of cross-reactive illness, such as GVHD triggered by CMV or other viral reactivation.

T cell cross reactivity originally was uncovered in relation to autoimmunity, especially in the context of CMV [12-15] and solid organ transplantation. [16] TCR-peptide-HLA class I bound complexes exhibit a strong recognition of 2-4 central amino acid residues in various orientations and allows for multiple amino acid substitution to the flanks in anchor positions while relying on HLA and peptide sequences simultaneously. The idea of T cells reacting to antigens with amino acid sequence homology (≥6/6 consecutive AA residues in a 9-mer for example) on different cell types with a given HLA-class I type has taken hold from multiple clinical examples. [13] Therefore, while CMV and other B-Herpesviruses have been shown to be associated with autoimmunity and contribute to oncogenic progression possibly affecting relapse rates, [17] CMV-specific T cell cross reactivity through adoptive lymphocyte transfer has been shown effective in the treatment of glioblastoma multiforme upon reactivation. [18,19] These clinical trials centering on confirmed CMV-specific T cell cross reactivity with infected tumor cells highlights the expectation of eliciting an immune response to CMV and 'self' tumor cells simultaneously. However, the development of CMV-specific T cells in that manner raises the question of eliciting a cross-reactive GVHD response against HLA bound-alloreactive recipient peptides. In theory, this will occur due to peptide polymorphisms, resulting from single nucleotide polymorphisms (SNPs), identified in the recipient that may mimic CMV peptides (*or vice versa*) with enough sequence homology to elicit strong HLA-specific immune responses, and presented as targets to donor lymphocytes (Figure 1).

Therefore, if CMV-derived HLA class I bound oligopeptide have sequence homology with human, alloreactive peptides bound to the same HLA class I and trigger cross recognition by donor T cells, the presence of these cross-reactive peptides may constitute a risk factor for the development of GVHD following SCT. In this study, HLA bound peptide-match analysis is facilitated by a newly compiled Human CMV CROSS (Cross-Reactivity Open Source Sequence) database. There were significant total CMV proteomic matches present, and in a majority of CMV sero-positive patients with GVHD, this was corroborated by GVHD-specific organ incidence of the human peptide source genes. These findings suggest that there may exist a potential CMV-Human polymorphic peptide influence or trigger, supporting a link between CMV reactivation and GVHD [8].

**Methods**

*Whole Exome Sequencing and Peptide Library Creation*

Patients with recurrent or high-risk hematological malignancies undergoing allogeneic SCT at Virginia Commonwealth University were included in this retrospective study following Institutional Review Board approval. Previously cryopreserved DNA samples, de-identified by clinical research staff, were whole exome sequenced. These recipients received an allograft from either a related or an unrelated, HLA-matched donor. Initially, 9 DRPs, including 4 HLA-matched related (MRD) and 5 unrelated (MUD) donor recipients, were annotated for identification of all the non-synonymous single nucleotide polymorphisms (nsSNPs), across the whole exome, in the graft versus host (GVH) direction (nsSNP(GVH)), i.e., polymorphism present in the recipient but absent in the donor. [10] Each nsSNP(GVH) was analyzed using the ANNOVAR software, [20] as previously described, [11] to populate the flanking amino acids around each amino acid coded for by each of the polymorphisms. This was accomplished through sequence padding, using DB SNP130 and hg18 genome coordinates, and yielded 17-mer peptides with variant amino acids occupying the central location. This 17-mer oligopeptide library created the opportunity to derive 9 separate 9-mer recipient oligopeptides, using a sliding window method, with the polymorphism-derived amino acid positioned at position 1-9 from C- to N-terminus. The nine oligopeptides generated per nsSNP(GVH) were extended to the whole exome for each recipient, representing each DRPs' unique peptide library. Subsequently, the number of DRPs analyzed was expanded to 77, with donor and recipient samples undergoing whole exome sequencing and peptide library generation.

*In Silico Peptide-HLA Binding Affinity Determination*

Patient peptide libraries were initially analyzed by the NetMHCpan software, version 2.8. [21,22] The analytic output yielded source gene information, polymorphic peptide sequence (9-mer), and a calculated IC50 value from the NetMHCpan algorithms for each of the six HLA class I molecules, HLA-A, B and C in each DRP. IC50 values (nM) indicated the amount of peptide required to displace 50% of intended or standard peptides specific to a given HLA. Binding affinity is inversely related to IC50 values such that a smaller IC50 value indicated a stronger affinity. The variant alloreactive peptides with a cutoff value of IC50 ≤500 nM to the relevant

HLA were included, and designated, *presented peptides*; and those that had an IC50 ≤50 nM, were termed *strongly presented* peptides. IC50 values in this range (IC50<500 nM) were considered relevant in previous studies of HLA-binding [23] and predictive mouse models of CTL response to vaccinia virus epitopes. [24] Further processing and interrogation of HLA binding in sliding 9-mer windows allowed for affinity sorting and HLA-specific separation along with gene information.

*Determining Sequence Homology Between Human And CMV Derived HLA Bound Peptides*

The bioinformatic pipeline utilized to interrogate SCT DRPs for HLA bound polymorphic peptides derived from nsSNP(GVH) in each DRP was extended and refined to interrogate each SCT DRP for relevant HLA class I bound CMV peptides. The binding of CMV-derived peptides to specific HLA Class I molecules was the first step towards determining their sequence homology with human HLA bound peptides. This bioinformatic pipeline is depicted in Figure 2. The initial step was *compilation* of the Human CMV-CROSS (Cross-Reactivity of Open Source Sequences). This is a database of 289 hCMV proteins or variants representative of the entire known multi-strain inclusive (Towne, Toledo and AD169) hCMV proteome sourced from NCBI (Appendix A; Protein variants had a single or multiple amino acid difference reported for the same protein/CMV gene; Deduplication was performed to remove bias). [25] The next step was the *utilization* of previously created patient polymorphic peptide libraries (derived from nsSNP(GVH) in each DRP) for initial feasibility of BLAST protein sequence alignment analysis. [26] This was followed by a subsequent confirmatory CMV-Human sequence homology analysis. Sequence homology relied on a match of 6 or more continuous amino acids in a string of 9 total amino acids (6/6-9/9) by sliding window analysis to identify sequence overlap between HLA bound 9-mer alloreactive human and hCMV peptides. Next, HLA Class I binding prediction (NetMHCpan) was performed for the hCMV peptides, screening the generated hCMV 9-mer peptides for HLA binding affinity across the test patient population (n=9).

A specific streamlined bioinformatics application (Table 1) was developed to directly compare the alloreactive donor-recipient peptide libraries for all the patients (n=77) to the HLA bound hCMV peptides by sequence homology; eliminating the need for the Protein BLAST sequence alignment (≥6/6 match by 9-mer sliding window analysis using all 9-mers extracted from each database, DRP peptide library or Human CMV CROSS proteome, pre- binding affinity screening).

Binding affinity was predicted (NetMHCpan) for all the 70,686 hCMV 9-mer peptides (resulting from the 289 hCMV proteins) to the known 2,915 human HLA-class I types along with collection of source gene information (hCMV and Human geneids), which were used to sort the peptides for strength of binding affinity. Immunogenicity screening was then performed to validate CMV-human matched peptides for the presence of hCMV gene products previously reported as CD8+ T cell targets known to elicit specific immune responses. [27]

*Tissue distribution of cross reactive peptides*

Homologous CMV-Human peptide libraries were compiled for each DRP and the source genes organized for putative GVHD target tissue-specific distribution analysis using the GTEx portal for expression data. [28] Each patient with GVHD involving specific tissues (e.g., skin) had tabulation of those disease-incident tissue-specific alloreactive peptides presented by their HLA (e.g., 10 skin peptides from 8 genes expressed in the skin) and the remainder GVHD tissue peptides/genes were tabulated as other or nonspecific to the patient's disease then tallied. Following criteria were used to assign CMV-Human peptide cross-reactivity potential: 1. Amino acid sequence homology of 6/6 to 9/9 (9-mers) between CMV and Human HLA bound peptides 2. IC50 values less Than 500 nM for binding to the unique HLA in each DRP (Presented or strongly presented peptides) 3. Human Gene-GVHD Tissue Expression Connection (Patient specific, i.e. skin, GI, etc.)

*Patients*

CMV reactivation defined as ≥250 copies of hCMV DNA/µL of plasma was determined by PCR. Absolute lymphocyte counts were measured at least twice weekly following SCT during the first 100 days and at least once weekly out to 6 months following SCT. Human CMV titers, lymphocyte counts and serum immunosuppressant drug levels were collected to one year post SCT from patient electronic medical records. Acute and chronic GVHD was diagnosed according to standard clinical criteria. Student's T test was utilized for statistical analysis.

**Results**

*Patient Characteristics*

Seventy-seven DRPs underwent exome sequencing following SCT, and were assessed by *in silico* analysis for alloreactivity potential derived from non-synonymous SNPs in the GVHD direction as previously described. [11] The cohort comprised of 27 MRD and 50 MUD SCT recipients underwent data collection for retrospective analysis including GVHD occurrence and hCMV reactivation information (Supplementary Table 1). Of the 30 patients that experienced hCMV viremia within a median of 29 days from transplant, 26 were hCMV-seropositive or reactivated; one patient had drug-refractory hCMV infection, and succumbed to it. CMV reactivation was experienced in conjunction with GVHD in 18 of the 30 patients, 13 of whom experienced hCMV reactivation prior to GVHD onset. Multiple recurrences of hCMV viremia were experienced in 11 of the 30 patients (2 or more separate reactivation events, ≥250 copies of hCMV DNA/μL blood per event).

*CMV - Human HLA bound peptide homology*

Based on the hypothesis that significant amino acid sequence homology between human- and viral-derived peptides presented on the same Class I HLA molecules may lead to donor CD8+ T cell cross reactivity, a bioinformatic pipeline was utilized to assess sequence homology between the patient's putative alloreactive peptide library, and hCMV-derived oligopeptides bound to each recipient's HLA. Whole exome sequencing of HLA matched SCT donors and recipients revealed 2,463 ± 603 nsSNPs (mean±SD) with a GVH direction per MRD and 4,287 ± 1154 nsSNPs per MUD recipient (Student's T-test, p<0.001). Following ANNOVAR 9-mer peptide determination and completion of the sliding window analysis of the resulting 17 amino acid variant oligopeptide resulting from each nsSNP, 43,705 ± 10,938 nonameric potentially alloreactive human peptides were identified per MRD recipient, and 77,025 ± 21,170 per MUD recipient. These were organized into a library by HLA-specific binding affinity (IC50: 0-50,000 nM). Next, each alloreactive Human 9-mer peptide library was further evaluated for their degree of match with the 70686 nonamer peptides derived from the hCMV proteome, utilizing the HLA-specific algorithmic CROSS database. This comparison determined sequence homology for strings of ≥6/6 consecutive amino acids between the two sets of Human and hCMV peptides, and was termed 'sliding window-match analysis'. This initial screen yielded an average of 29,659

± 9039 total peptide matches per MRD patient and 52,910 ± 16122 total peptide matches per MUD patient to the hCMV proteome (Student's T-test, p<0.001) following analysis. The program output reported for each peptide match, the geneid's, the HLA, IC50 values (Range: 0-50,000 nM) and complete peptide sequence, as well as the shared partial peptide for degree of sequence homology.

Upon confirmation that all 77 patient alloreactive peptide-HLA libraries studied had matches with the hCMV peptide-HLA arrays, the degree of sequence homology in tightly HLA bound peptides (IC50: 0.01-500 nM) was determined. Following removal of duplicate peptide sequences, the total CMV-human match (homology) library, yielded an average of 33 peptide matches per MRD patient and 44 peptide matches per MUD patient (Student's T-test, p=0.09), constituting approximately 1% of the total matches reported (Table 2). The homologous sequence information per patient's peptide library following high affinity match analysis (IC50<500) was compiled into Appendix B; this includes an immunogenicity sorting performed for hCMV reactivating patients. Considering the degree of sequence homology present between hCMV and human peptides bound to HLA class I molecules in specific DRP, each MRD DRP on average had 31, 2, 0 and 0 relevant nonameric CMV-Human peptide matches with 6/6, 7/7, 8/8 and 9/9 sequence homology respectively (Table 2). Also each MUD DRP on average had 40, 3, 1, and 0 peptide matches with 6/6 - 9/9 sequence homology respectively. As expected from the donor type SNP differences, relevant alloreactive peptide differences, and total peptide matches with hCMV, MUD DRPs exhibited a trend for a higher number of hCMV matched peptides than MRD DRPs by peptide sequence homology differences (6/6-9/9 matching; student's T-test, p=0.07-0.38 respectively). To determine the correlation of the number of high affinity CMV-Human matches (IC50<500 nM) with sequence homology >6/6 amino acids in 76/77 patients, these were plotted against the pool of total alloreactive peptides per patient (Figure 3). Notably, MUD DRPs post screening (IC50<500 nM) exhibited a mean of 6,545 ± 2689 alloreactive peptides per patient which was significantly greater than the 4,522 ± 1915 mean alloreactive peptides per MRD DRPs (Student's T-test, p<0.001) prior to interrogating for matches to the hCMV proteome, with a trend for a higher prevalence of CMV-Human matches in the MUD recipients. These results indicate that there exists a pool of homologous-CMV-derived peptides, which may be presented by the recipient HLA and trigger cross-reactive donor T cell identification of recipient alloreactive peptides; therefore hCMV reactivation may potentially trigger GVHD in some patients. These data also identify MUD recipients for being at a somewhat higher risk of possible alloreactivity resulting from hCMV reactivation.

*Potential Immunogenecity of the hCMV Peptides Presented on Recipient HLA*

To compare the sequence homology analysis results and determine the presence of previously validated immunogenic hCMV peptides, [27] the specific hCMV peptide sequences and the proteins of origin were reviewed (Table 3). Twelve of the 13 patients with hCMV reactivation before GVHD onset exhibited one or more immunogenic hCMV peptide matches, previously shown to elicit a hCMV-specific T-cell response targeting the listed source genes. The remaining patient expressed CMV – Human peptide-matches specific to the patient's affected GVHD tissue, which had not previously been reported in other studies. The analysis revealed multiple CMV-Human peptide matches that were high affinity (IC50<500 nM), immunogenic and expressed in a GVHD-affected tissue-specific manner. Further, the binding affinity (reflected by the IC50 values) of hCMV peptides when plotted against the IC50 values of the alloreactive Human peptides (Figure 4), demonstrated a wide range of values, indicating the potential for varying degrees of cross reactivity. This analysis illustrates the magnitude of overlap in peptide sequence between the two sources of HLA presented peptides and provides a rational explanation for cross reactivity potentially triggering GVHD.

*Tissue expression of cross-reactive Human peptides*

In order to study the GVHD tissue-specific expression of the hCMV peptides matched to human alloreactive peptides (Table 3), gene expression data organized in a tissue specific manner was obtained from the GTEx Portal of the Broad Institute of MIT and Harvard (Version 6) [28]. Gene expression data for all CMV-Human peptide matches in the 30 hCMV reactivating/*de novo* infected patients was compiled (median minimum threshold of expression: ≥10 reads/kilobase of transcript/million mapped reads), specifically focusing on GVHD target tissues including: skin, GI, liver, lung, and others (vagina, muscle, adipose and salivary gland). A focused analysis of GVHD tissue gene expression in conjunction with actual GVHD occurrence was performed on 18 patients with GVHD and hCMV reactivation, and showed that 18/18 patients had expression of hCMV-matched alloreactive peptides (IC50<500 nM) with ≥10 RPKM tissue-specific gene expression (Table 4). Combining both match data and gene expression data, a cross reactivity profile was created for the hCMV infected subset of patients to include: match number, discrete numbers of peptides (Human and CMV), gene count, immunogenic hCMV genes, and tissue specific GVHD peptides or genes. These data comprise a predictive case for potential alloreactive trigger following hCMV infection that can be derived, *a priori*, from whole exome sequencing of transplant donors and recipients using this analytic approach.

*Clinical correlations*

In this cohort hCMV reactivation following SCT, was associated with poorer survival (Log Rank: 4.5, p=0.03, Hazard ratio: 2.31, p=0.04) when adjusting for hCMV reactivation by donor or recipient seropositivity (Supplemental Figure 1), consistent with the findings of a recent CIBMTR analysis [2]. To determine whether there was any association between onset of hCMV reactivation and GVHD, we evaluated lymphocyte growth kinetics and large variations in calcineurin inhibitor levels as possible triggers of GVHD occurrence in the above referenced patients (Table 4). It has been demonstrated that lymphocyte recovery after SCT, follows logistic dynamics, with departures from the logistic curve being recorded at the time of infection/relapse etc.[29] By examining case studies of multiple hCMV-reactivating patients, we observed that their reactivation events were at times coincident with GVHD onset and often associated with preceding lymphocytosis, suggestive of an alloreactive cellular immune response. Acute GVHD onset following CMV reactivation was present in 7/13 hCMV reactivation before GVHD patients; acute GVHD alone without chronic GVHD indicated poorer outcome in this cohort of patients, with 3 patients experiencing Grade IV GI disease (two were steroid refractory) which was fatal in all cases. As seen in figure 5A, with acute GVHD in patients 47, 68, 70, there was evidence of hCMV reactivation events (top graph) that preceded a burst of lymphocytosis (middle graph) and often occurred with stable CNI or immunosuppression levels (bottom graph), indicating a potential hCMV-GVHD relationship. Patients 47, 68 and 70 with Acute GI GVHD had an average of 20.3 ± 9.3 matches (±SD) derived from 18.7 ± 7.8 human peptides. Patient 68 also exhibited skin and liver GVHD. When comparing the Acute GVHD patients to Chronic or Acute + Chronic, we did note a 'quality over quantity' trend emerging when considering the percentage of genes with GVHD-specific tissue expression on average, in these 3 patients GI tract specific alloreactive-hCMV matched genes were 80.6 ± 17.3 %. Patients depicted in figure 5B exhibited the more stable or gradual hCMV reactivation effects seen in chronic GVHD patients 10 and 27, where lower or less frequent spikes in hCMV titers may still elicit lymphocyte growth just at a more controlled rate, probably accounting for that perceived difference in outcomes between patients with symptoms of only acute GVHD or chronic GVHD. Patients 10 and 27 with Chronic Skin GVHD had a mean of 68 ± 5.7 matches (±SD) derived from 50.5 ± 10.6 human peptides. These two patients exhibited a lower percentage of GVHD-tissue specific gene expression on average when considering all potential GVHD genes identified, in this instance skin gene expression amounted to 51.9 ± 2.7%. The final two cases (patients 71 and 84, Figure 5C) exhibited symptoms of both acute and chronic GVHD, but patient 84's lower grade cyclical reactivation events altered the lymphocyte growth pattern towards a sinusoidal pattern. Patients 71 and 84 exhibiting both acute and chronic GVHD had a mean of 40 ± 21.2 matches (±SD) derived from 25 ± 5.7 human peptides. Patient 84 also exhibited GI GVHD. Patient 71 was *de novo* infected and continued with a low-grade infection for 200+ days prior to declining counts towards relapse. Patients 71 and 84 also exhibited lower GVHD-tissue specific gene expression on average when considering all potential GVHD genes identified, Skin or GI (primary organs): 45.8 ± 29.5%. These variable dynamics of hCMV reactivation and GVHD onset demonstrate the complexity involved in analyzing the relationship between ongoing immunosuppression in the setting of multiple sets of potentially cross reactive antigens by affected organ system being presented to a reconstituting donor-derived immune system.

**Discussion**

CMV reactivation is a frequent complication of allografting, requiring frequent monitoring and associated with an increased risk of treatment related mortality, primarily in its own right, but also because it is frequently associated with GVHD. [8] Therapy and effective prophylaxis involve the use of toxic drugs and monitoring for reactivation is not straightforward. The ability to identify patients at risk of developing alloreactive complications from hCMV reactivation will therefore be a useful adjunct to the supportive care of transplant recipients, as well as an important step forward in understanding virus induced alloreactivity. In this paper a computational algorithm that identifies hCMV peptides homologous to human alloreactive peptides is described. This determination required three steps, whole exome sequencing of transplant donors and recipients, followed by *in silico* determination of the patient specific class I HLA binding of the oligopeptides resulting from the nsSNP in the exome and finally a comparison of these alloreactive peptide sequences with those of hCMV peptides predicted to bind the same HLA molecules. This algorithm identifies a number of hCMV peptides which bind the same HLA as human alloreactive peptides with a similar range of binding affinities and may potentially be cross presented to donor T cells.

To understand how this may impact GVHD pathophysiology, consider a T cell clone (TC$_{CMV}$) which recognizes these hCMV peptide-HLA complexes, is activated by a hCMV viremia. The T cell receptor of this clone may also recognize a human alloreactive peptide with sequence homology to the hCMV peptide and bound to the same HLA molecule. Even if it does so weakly, tissue damage may be initiated and GVHD ensue. This process can work in reverse as well, a T cell clone with high affinity for alloreactive peptides (TC$_{mHA}$), which only binds the hCMV peptide-HLA complex weakly may be 'set-off' by a hCMV reactivation event, again leading to down stream GVHD. This general principle likely holds true for most viruses.

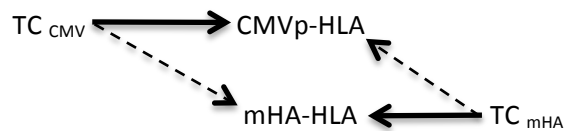

A mathematical model utilizing matrices has been developed to understand aggregate T cell responses to many mHA-HLA complexes the donor T cells may encounter in the recipient milieu. [30] The simplifying notion underlying this model is that each T cell interacts with a single antigen, therefore an identity matrix may be used to calculate the resulting T cell response. This model requires that the antigens occupy the matrix operator ($M_{APO}$) and T cell vector gets transformed by the operator, as seen in the matrix below.

| SCT | | $\mathbb{M}_{APO}$ | | | |
|---|---|---|---|---|---|
| | | mHA$_1$ HLA | mHA$_2$ HLA | mHA$_3$ HLA | mHA$_n$ HLA |
| $\overline{v_{TCD}}$ | TC$_1$ | 1 | 0 | 0 | 0 |
| ↓ t | TC$_2$ | 0 | 1 | 0 | 0 |
| $\overline{v_{TCR}}$ | TC$_3$ | 0 | 0 | 1 | 0 |
| | TC$_n$ | 0 | 0 | 0 | 1 |

In the above simplified matrix T cell clone, TC$_1$ only interacts with mHA$_1$, and so on. The 1 in the cells means that the T cells recognize that antigen and responds, and 0 means absence of recognition. In reality, the TCR-Ag-HLA interactions are not likely to be quite so simple. An important clue to this is the observation that antigen-HLA binding as indicated by IC50 does not occupy discrete states of 0 or 1 (binding or no binding); instead there is a continuum of IC50 values. So in the case of cross-reactive antigens, a hCMV derived antigen binds the same HLA molecule and interacts with the T cell receptor, albeit with a different binding affinity. This implies that each TCR might interact with multiple antigens with different affinity, thus the 0 gets replaced with a series of numbers between 0 and 1. These cross-reactive antigens triggered by weakly interacting antigens may augment the T cell response to the target antigens.

| SCT | | $\mathbb{M}_{APO}$ | | | |
|---|---|---|---|---|---|
| | | mHA$_1$ HLA (CMVp$_1$ HLA) | mHA$_2$ HLA (CMVp$_2$ HLA) | mHA$_3$ HLA (CMVp$_3$ HLA) | mHA$_n$ HLA (CMVp$_n$ HLA) |
| $\overline{v_{TCD}}$ | TC$_1$ | 1 | (0.2*) | 0 | 0 |
| ↓ t | TC$_2$ | 0 | 1 | (0.1*) | 0 |
| $\overline{v_{TCR}}$ | TC$_3$ | (0.3*) | 0 | 1 | (0.1*) |
| | TC$_n$ | 0 | 0 | (0.5*) | 1 |

\* indicates response of the relevant T cell clone to a viral pathogen peptides, bound to the same HLA as the mHA and recognized by the same T cell clone.

This matrix demonstrates the perspective from an alloreactivity operator, however if one were to think about the operator constituting a hCMV infected tissue, the following distribution of T cell interactions may be observed

| SCT | $\mathbb{M}_{APO}$ | | | | |
|---|---|---|---|---|---|
| | | CMVp$_1$ HLA (mHA$_1$ HLA) | CMVp$_2$ HLA (mHA$_2$ HLA) | CMVp$_3$ HLA (mHA$_3$ HLA) | CMVp$_n$ HLA (mHA$_n$ HLA) |
| $\overline{v_{TCD}}$ $\downarrow t$ $\overline{v_{TCR}}$ | TC$_{1CMV}$ | (0.3*) | 1 | 0 | 0 |
| | TC$_{2CMV}$ | 0 | (0.2*) | (0.1*) | 0 |
| | TC$_{3CMV}$ | 1 | 0 | (0.5*) | 1 |
| | TC$_{nCMV}$ | 0 | 0 | 1 | (0.1*) |

\* indicates alloreactive mHA, bound to the same HLA as the viral peptides and recognized by the a different set of T cell clones, which recognize the hCMV peptide as the primary target and have the mHA as a target with cross reactivity.

Above conceptual explanation aside, clinical correlation indicates that patients meeting the previously described criteria for cross reactivity, experienced hCMV reactivation and subsequent GVHD onset, as noted in the sevral case studies presented. It can be seen that hCMV reactivation directly preceding lymphocytosis was found in a subset of our population which may indicate an autoimmune like relationship based on genetic predisposition or the quality of the matching peptides bound with great affinity to specific HLA with appropriate protein abundance or gene-tissue expression for GVHD to occur.

While the *in silico* analysis for hCMV + alloreactivity potential reveals a large body of antigens which may influence clinical outcomes, there are caveats to be considered in developing this peptide analysis pipeline. First, the process of creating the CMV CROSS database included some variant forms of the hCMV proteins known to have differing amino acid lengths that may allow for duplicates, which were accounted for during multiple processes of deduplication. However the slight differences to the variant forms of the proteins reported in the NCBI database may generate a few more peptide possibilities than may be realistic but would allow for strain differences of the human peptide matches during exome sequencing. The actual process of matching peptides or component sequences (6/6-9/9) also allowed for multiple match possibilities by slight human 9-mer sliding window changes to peptides, multiple matches to the same CMV proteins at different sites, strings of repeated amino acids (i.e., AAAAAA or SSSSSS) allowing for multiple matches or limited peptide match possibilities from terminal amino acid polymorphisms on various proteins. These points may introduce some small bias, which was accounted for by removing duplicate matches at multiple time points and counting genes or peptides discreetly apart from the number of matches, which may be inflated in the described instances above. An interesting note in this same vein was observed among patient or hCMV peptides, where human genes and hCMV immunogenic genes were shared among multiple patients with many of each gene or peptide involved being common in our cohort, often with the same HLA specificity and predicted binding affinity, resolving that there is likely still a more complex pattern to unravel from this approach to cross-reactivity potential (Appendix B).

Epitope spreading is another phenomenon, in stem cell transplant similar to autoimmune disorders, that would be exemplified by donor lymphocytes and antigen presenting cells encountering either hCMV or recipient alloreactive peptides released upon organ damage that may be processed and presented on restricted HLA with enough sequence homology to their ligands, hCMV or alloreactive peptides, to elicit an immune response. Also an inherent genomic similarity might be expected over a long course of coevolution due to viral latency where hCMV has adopted our genetic information for immune evasion in the majority of the population by molecular mimicry.

CMV has evolved in parallel with the human genome over hundreds of millions of years, potentially exchanging genetic information from virus and human randomly with each latent infection, selecting primarily for immune evasion. Human CMV has adopted many well-known immune evasive strategies but employing human peptides for the purpose of molecular mimicry, including the production of decoy HLA-peptide complexes or immunosuppressive viral IL10, has proven to be one of the most successful strategies. [31] This molecular mimicry shields hCMV from a portion of the immune response but also exposes endogenous antigens from recipient cells simultaneously with hCMV antigens to donor immune surveillance during organ damage from lytic reactivation in athymic adults. [32] This form of epitope spreading, developing T cell antigenic experience, affects overall immune responses and may account for the robust CD8+ T cell response to hCMV infection in otherwise healthy individuals (10% of the entire T cell compartment) that inflates with age. [27] In addition, the primarily memory T cell response to hCMV accounts for the sharp responses to various hCMV antigens upon reactivation, which may be as great as 50 fold the strength of a naïve T cell response during primary infection. [15] Prior evidence of eliciting a strong CD8+ and CD4+ response to soluble recombinant hCMV antigen, measured by IFN-gamma production, [33] indicates the presence of epitope spreading in the realm of post-transplant viral immunity as it has already been discovered in Multiple Sclerosis and implicated in other autoimmune disorders. [17,34,35]

When considering the opportunity for hCMV to catalytically trigger GVHD apart from the potential method we propose here, we must first note that hCMV is the largest of the herpesvirus family with a 235 kb DNA genome capable of encoding more than 200 potential protein products [36] and second the near ubiquity of cells it is capable of infecting in man, including parenchymal and connective tissue cells of virtually any organ along with various hematopoietic cell types (epithelial, endothelial - including vascular and each organ, fibroblasts, smooth muscle cells, monocytes and macrophages). [37] Most chronic human viral genomes experienced in the SCT population are much smaller in size, e.g. adenovirus 26-45 kb, [38] BK virus and JC virus are approximately 5 kb [39] and most respiratory viruses <20 kb, [40,41] indicating a much greater potential peptide pool for hCMV to draw from for purposes of molecular mimicry and immune evasion. The potential for hCMV infection in nearly any cell type obviates the noted wide scale CD8+ T cell expansion that occurs upon reactivation but also the potential for epitope spreading upon organ damage throughout the human body, especially the large surface area tissues with immune exposure like the vasculature, skin, gastrointestinal tract, liver and lungs. When we compare hCMV to the other herpesviruses on tissue specificity, HHV6 infects a slightly more restricted variety of human cells but prefers replication *in vitro* in CD4+ T cells which HHV7 infects primarily, [42], EBV infects B cells or epithelial cells [43] and HSV (150 kb) can infect many cells while its latency is neuron restricted. [44] Thus, hCMV seems uniquely well suited to have a high potential for molecular mimicry as well as its presence in the vasculature and epithelium of almost every organ.

In conclusion, this paper reports sequence homology in HLA bound peptide antigens of hCMV and human origin. Given the distribution of human peptides in various tissues, and their involvement with GVHD in the patients examined, we posit that hCMV derived peptides may influence the development of GVHD in patients who develop hCMV reactivation following SCT. These findings support the use of more aggressive antiviral strategies for preventing hCMV reactivation in patients undergoing allografting and argue against the use of simply monitoring as the major therapeutic strategy. We endeavor in the future to understand the supporting CD4+ T cell hCMV protective dynamics involved by interrogating HLA class II peptides, [1] looking at potential influences of human CMV mimicry and unraveling the potential GVHD/CMV cross-reactivity/Auto-immune relationship further [9,14,17,45-49].

**Acknowledgements**

This study conducted at Virginia Commonwealth University's Massey Cancer Center was supported, in part, by research funding from the NIH-NCI Cancer Center Support Grant (P30-CA016059) to Amir Toor and by research funding from Virginia's Commonwealth Health Research Board Grant #236-11-13 to Michael Neale. Sequencing and Bioinformatics Analysis was performed in the Genomics Core of the Nucleic Acids Research Facilities at VCU. We appreciate the critical review of our work provided by Jeremy Meier, VCU student MD-PhD program, and all others who supported this work in some way.

**Tables.**

**Table 1**. Match Algorithm & Analytic Design.

| Table 1. | HCMV cross-reactivity potential of 77 SCT DRP from whole exome sequencing data using sliding window match analysis |
|---|---|
| 1 | 289 HCMV protein sequences representative of the HCMV proteome were downloaded from NCBI by batchentrez utility |
| 2 | Downloaded CMV protein sequences were sliced into 9AA peptides and deduplicated, resulting in a library of 70686 9AA peptides |
| 3 | Each of the 70686 9AA peptides were further divided into 2 8AA, 3 7AA and 4 6AA peptides to form a lookup table for each component peptide (6,7,8) with the one or more 9AA parent peptide(s) and the associated CMV geneid |
| 4 | As a focus on Class I HLA, a list of the 2915 plausible HLA-A/B/C types interrogated by netmhcpan was compiled |
| 5 | IC50 values for all 70686 9AA CMV peptides were precalculated against all 2915 HLA types and saved in the CROSS database |
| 6 | 9-mer peptides isolated from ANNOVAR per patient were searched for match with the 9AA CMV peptide pool<br>… if no match found then each 9-mer was split into 2 8-mers and each searched in the 8AA CMV peptide pool<br>… if still no match found then each 9-mer was split into 3 7-mers and each searched in the 7AA CMV peptide pool<br>… if still no match found then each 9-mer was split into 4 6-mers and each searched in the 6AA CMV peptide pool |
| 7 | If any of the above searches found a match, the search was terminated and results for that patient 9-mer peptide reported |
| 8 | Reported 9-mers were linked with geneid (Human & CMV), HLA type, IC50 (Human & CMV), partial peptide (6/7/8), CMV_peptide (9) |
| 9 | Each patient-specific peptide library generated by comparing alloreactive peptides with the CROSS database were compiled, statistically analyzed (Student's T-Test) and further screened for high affinity binding (1/IC50 or an IC50 range of 0-500 nM) |
| 10 | Binding strength (IC50<500 nM) of homologous CMV and human peptides to DRP-specific HLA was compared for pattern analysis |
| 11 | Immunogenicity of common CMV gene products as CD8+ T cell targets was screened for validation on CMV infected patients (n=30) |
| 12 | CMV peptide matches to Human alloreactive peptides underwent multiple rounds of deduplication to ensure only unique peptide inclusion by both software programming (Excel or independent coding) and finally by human observation upon final analysis. |
| 13 | Gene Tissue Expression (GTEx) data was gathered on CMV infected patients (n=30) to find GVHD specific organ incidence |
| 14 | Case studies were prepared, 7 patients with CMV before GVHD: 3 acute GVHD, 2 chronic GVHD and 2 aGVHD + cGVHD. |

**Table 2.** CMV- Human match analysis per DRP (IC50 <500 nM) by donor type according to continuous amino acid sequence homology (one standard deviation from the mean value)

| Mean Relevant Matches (IC50<500) | Number of Patients (n) | Sequence Homology 6/6 (±SD) | Sequence Homology 7/7 (±SD) | Sequence Homology 8/8 (±SD) | Sequence Homology 9/9 (±SD) | Total 6/6 to 9/9 (±SD) |
|---|---|---|---|---|---|---|
| MRD | 27 | 31 ± 19 | 2 ± 2 | 0 ± 1 | 0 ± 0 | 33 ± 20 |
| MUD | 50 | 40 ± 33 | 3 ± 6 | 1 ± 2 | 0 ± 1 | 44 ± 39 |
| Overall | 77 | 37 ± 29 | 3 ± 5 | 1 ± 2 | 0 ± 1 | 40 ± 34 |
| **Student's T-test (p<.05)** | | 0.12 | 0.09 | 0.07 | 0.38 | 0.09 |

Table 3. GVHD tissue-specific immunogenic CMV peptide matches, CMV reactivation before GVHD patients. Patient-specific process of cross comparison along with the source genes, IC50 values to indicate inverse binding affinity and GVHD organ specific tissue involvement.

| Patient | HLA | Human Gene | IC50 Value | DRP Peptide | CMV Peptide | IC50 Value | CMV Gene | GVHD Organ with Target Gene Expression |
|---|---|---|---|---|---|---|---|---|
| 10 | HLA-B40:01 | TSPYL1 | 11.67 | SETVALPPL | QENSETVAL | 23.37 | UL86 | Skin |
| | HLA-B40:01 | HDAC7 | 26.71 | EEVEAVTAL | AEREAVTAL | 19.95 | UL49 | |
| | HLA-B08:01 | APEH | 486.52 | NRRSALYSV | MWKNRRSAL | 10.53 | UL37 | |
| | HLA-B08:01 | | | | WKNRRSALL | 273.97 | UL37 | |
| 27 | HLA-C03:04 | IL22RA1 | 15.86 | IVHPTPTPL | SSTPTPTPL | 41.56 | UL82 | Skin |
| | HLA-A02:01 | TECR | 28.47 | ALFSLVVFT | YLFSLVVLV | 2.76 | US3 | |
| | HLA-A02:01 | | | | LVYLFSLVV | 441.74 | US3 | |
| | HLA-C03:04 | WNK1 | 41.43 | ITAAATAPV | TTAAATATV | 198.72 | UL105 | |
| 45 | HLA-A68:02 | PPP1R15A | 149.06 | SSAAAAAAL | TAAAAAAAA | 57.77 | IRS1,TRS1 | Skin & Vagina |
| | HLA-A68:02 | | | | ATAAAAAAA | 59.30 | IRS1,TRS1 | |
| | HLA-A68:02 | | | | DAAAAAAPT | 285.57 | IRS1 | |
| | HLA-A68:02 | | | | SVSSSAAAA | 376.07 | UL105 | |
| | HLA-A68:02 | | | | SASAAAAAA | 468.89 | UL105 | |
| | HLA-A68:02 | | | | DATAAAAAA | 469.93 | IRS1,TRS1 | |
| | HLA-A68:02 | MUC21 | 106.17 | ESSTTSSGA | TTYTTSSGA | 23.86 | UL86 | Vagina |
| | HLA-A68:02 | | | | YTTSSGAKI | 139.73 | UL86 | |
| | HLA-A68:02 | HOXD13 | 351.91 | AAAAAAAGA | TAAAAAAAA | 57.77 | IRS1,TRS1 | |
| | HLA-A68:02 | | | | ATAAAAAAA | 59.30 | IRS1,TRS1 | |
| | HLA-A68:02 | | | | DAAAAAAPT | 285.57 | IRS1 | |
| | HLA-A68:02 | | | | SASAAAAAA | 468.89 | UL105 | |
| | HLA-A68:02 | | | | DATAAAAAA | 469.93 | IRS1,TRS1 | |
| 47 | HLA-B07:05 | MUC17 | 52.25 | TPVSHTLVA | TPVSHTQPL | 7.58 | UL150 | GI |
| | HLA-A30:02 | CARS2 | 342.62 | SPASLSSLY | YFDSLSSLY | 173.23 | UL29,UL28 | |
| 48 | HLA-A31:01 | PLXND1 | 200.40 | LFVFCTKSR | NLFVFCTER | 57.60 | UL34 | GI (Skin & Liver) |
| | HLA-A31:01 | PLXND1 | 307.31 | VALFVFCTK | | | | |
| 67 | HLA-A02:01 | FGFR4 | 19.97 | RLLLALLGV | VLLLALLLL | 208.77 | US29 | GI, Lung (Skin) |
| | HLA-A02:01 | FGFR4 | 164.97 | LLLALLGVL | | | | |
| 68 | HLA-A31:01 | ALCAM | 187.60 | TYTLTAVRR | ASLLTAVRR | 68.25 | IRS1,TRS1 | GI, Liver (Skin) |
| 70 | HLA-C07:02 | LCN2 | 121.85 | SYPGLPSYL | MRPGLPSYL | 49.96 | UL75 | GI |
| | HLA-B35:01 | SERPINB4 | 97.29 | EAAAATAVL | CAAAAATAA | 393.01 | UL105 | |
| | HLA-A03:01 | FYCO1 | 488.05 | RASLKRLVK | STLKRLVKK | 83.46 | UL69 | |
| 71 | HLA-C03:03 | PPP1R15A | 19.38 | SSAAAAAAL | AAAAAAPTV | 48.37 | IRS1,TRS1 | Skin |
| | HLA-C03:03 | | | | AAAAAAAAA | 343.39 | IRS1,TRS1 | |
| | HLA-C03:03 | | | | TAAAAAAAA | 388.91 | IRS1,TRS1 | |
| | HLA-C03:03 | | | | SASAAAAAA | 421.08 | UL105 | |
| | HLA-A30:02 | CARS2 | 342.62 | SPASLSSLY | YFDSLSSLY | 173.23 | UL29,UL28 | |
| 76 | HLA-A03:01 | ZZEF1 | 119.19 | SVLSELLKK | SVLSELLNK | 90.92 | UL54 | GI |
| | HLA-B27:05 | MUC4 | 414.06 | TRHATSLPV | YQLRHATSL | 473.18 | US29 | |
| 90 | HLA-A02:01 | PELP1 | 77.94 | LLALLLAPT | LLALLLLEL | 37.59 | US29 | Skin & GI |
| | HLA-A02:01 | | | | VLLLALLLL | 208.77 | US29 | |
| | HLA-A02:01 | AHSA1 | 321.16 | KTLFLAVQV | ELFLAVQFV | 144.94 | UL86 | |
| | HLA-C03:03 | TCF7L1 | 106.99 | AAASSSGQM | AAASSSSAV | 26.73 | UL48 | GI |
| | HLA-C03:03 | | | | IAAASSSSA | 219.78 | UL48 | |
| | HLA-A03:02 | ZZEF1 | 325.00 | SVLSELLKK | SVLSELLNK | 184.87 | UL54 | |
| | HLA-C03:03 | PLEKHA6 | 384.99 | AASSSLRRL | FGGAASSSL | 71.10 | UL122 | |

Note: Twelve patients with CMV reactivation/infection before GVHD onset exhibited previously identified immunogenic CMV peptide matches with gene expression specific to the tissues affected by GVHD (above); The filter of immunogenicity provides a connection to T cell reactivity shown in vitro to the listed CMV genes in a separate patient population; Patient 79 with muscle/fascia GVHD showed no muscle-specific previously *known immunogenic* CMV peptide matches but still had three relevant CMV peptide matches expressed in the skeletal muscle (not shown); Tissues in parentheses were also affected by GVHD but without immunogenic matches/expression by patient; Patients 67 and 71 experienced *de novo* CMV infection.

**Table 4.** Human-CMV short sequence homology in GVHD Tissue Specific Peptide and Gene Distribution from GTEx Analysis (n=18). GVHD incidence denotes the specific organs affected in each patient; Peptides, lists the number of unique peptide-HLA complexes matched between human and CMV peptide library; the column, *Genes* lists the source genes for the afore mentioned peptides; *GVHD tissue specific peptides* lists the number of peptides which bind HLA with an IC50 <500nM, and are expressed in tissues affected by GVHD; *GVHD tissue gene expression* denotes the number of genes expressed at an RPKM >10 corresponding to the GVHD tissue specific peptides.

| Patients | GVHD Incidence | | | | | | Peptides | | | Genes | | | GVHD Tissue Specific Peptides | | | | | | GVHD Tissue Specific Gene Expression | | | | | |
|---|---|---|---|---|---|---|---|---|---|---|---|---|---|---|---|---|---|---|---|---|---|---|---|---|
| n=18 | Skin | GI | Liver | Lung | Vaginal | Muscle | Human | Matches** | CMV | Human | Immunogenic CMV | CMV | Skin | GI | Liver | Lung | Vaginal | Muscle | Skin | GI | Liver | Lung | Vaginal | Muscle |
| 10* | X | | | | | | 58 | 72 | 58 | 47 | 10 | 35 | 10 | | | | | | 8 | | | | | |
| 27* | X | | | | | | 43 | 64 | 55 | 39 | 9 | 35 | 7 | | | | | | 7 | | | | | |
| 45* | X | | | | X | | 17 | 45 | 26 | 15 | 5 | 15 | 7 | | | | 9 | | 6 | | | | 8 | |
| 47* | | X | | | | | 21 | 23 | 22 | 20 | 5 | 19 | | 6 | | | | | | 6 | | | | |
| 48* | X | X | X | | | | 24 | 25 | 21 | 21 | 5 | 18 | 5 | 11 | 5 | | | | 4 | 9 | 4 | | | |
| 67* | X | X | | X | | | 28 | 43 | 36 | 24 | 4 | 24 | 3 | 8 | | 9 | | | 3 | 7 | | 7 | | |
| 68* | X | X | X | | | | 10 | 10 | 10 | 10 | 2 | 10 | 2 | 7 | 2 | | | | 2 | 7 | 2 | | | |
| 70* | | X | | | | | 25 | 28 | 22 | 21 | 5 | 21 | | 9 | | | | | | 8 | | | | |
| 71* | X | | | | | | 29 | 55 | 39 | 28 | 8 | 30 | 2 | | | | | | 2 | | | | | |
| 76* | | X | | | | | 10 | 11 | 11 | 10 | 4 | 10 | | 5 | | | | | | 5 | | | | |
| 79* | | | | | | X | 39 | 49 | 41 | 28 | 7 | 26 | | | | | | 3 | | | | | | 2 |
| 84* | X | X | | | | | 21 | 25 | 21 | 18 | 4 | 14 | 1 | 7 | | | | | 1 | 6 | | | | |
| 90* | X | X | | | | | 39 | 56 | 43 | 36 | 7 | 27 | 8 | 17 | | | | | 7 | 16 | | | | |
| 34 | X | | | | | | 38 | 57 | 47 | 32 | 12 | 31 | 8 | | | | | | 7 | | | | | |
| 38 | X | | | | | | 32 | 47 | 41 | 30 | 6 | 29 | 6 | | | | | | 6 | | | | | |
| 41 | | X | | | | | 12 | 12 | 11 | 11 | 2 | 11 | | 5 | | | | | | 4 | | | | |
| 73 | X | X | | | | | 5 | 5 | 5 | 5 | 0 | 4 | 1 | 2 | | | | | 1 | 2 | | | | |
| 81 | | X | | | | | 10 | 17 | 10 | 6 | 0 | 6 | | 4 | | | | | | 2 | | | | |

Note: *- All patients with an asterisk following their numeric representation experienced CMV reactivation prior to GVHD (except Patients 67 and 71, *de novo* CMV infected) and patients without an asterisk experienced GVHD prior to CMV reactivation.
**- human peptides may have overlapping areas of homology yielding a higher number of matches

**Figures.**

**Figure 1.** Proposed cross-reactivity model for CMV-specific T cells that may react to matched homologous peptides (≥6/6 consecutive amino acids) bound to the same HLA on normal recipient cells present in GVHD-affected tissues.

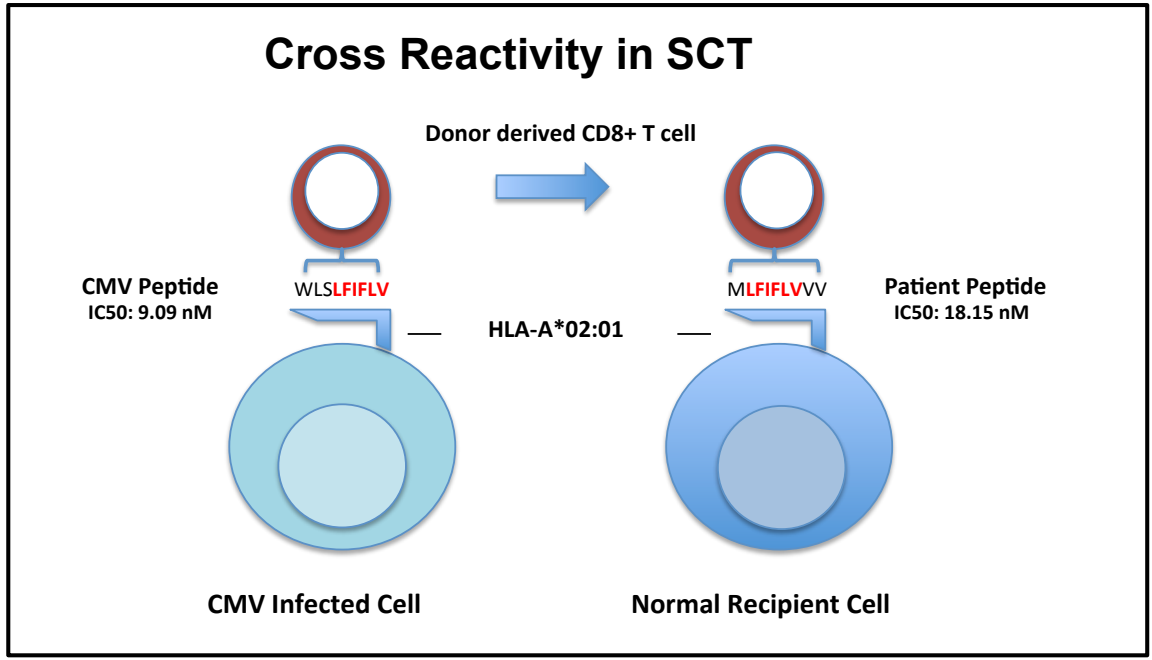

**Figure 2.** Bioinformatic pipeline comparing alloreactive peptide libraries with the CMV proteome for determination of potential Pro-GVHD DRP cross reactivity upon CMV reactivation correlated with GVHD-tissue specific distribution of peptides.

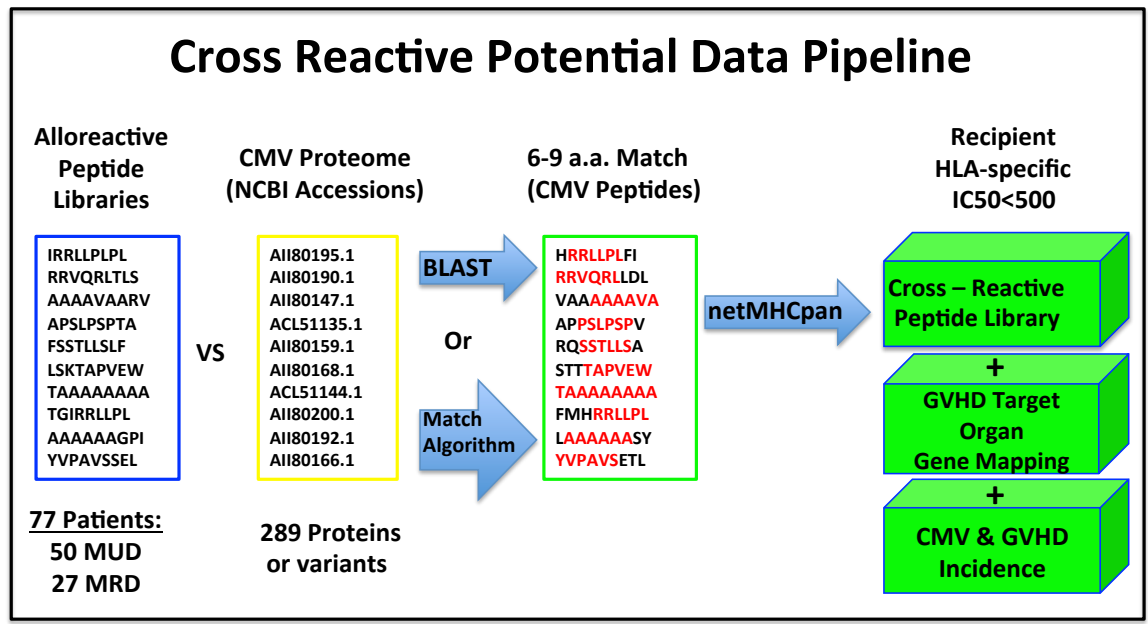

**Figure 3.** CMV+GVHD Cross Reactivity Potential: Patient-specific peak CMV peptide matches intersecting peak alloreactive peptides (IC50<500 nM) as a cross-reactivity potential stratified by donor type contained within each DRP alloreactive peptide library.

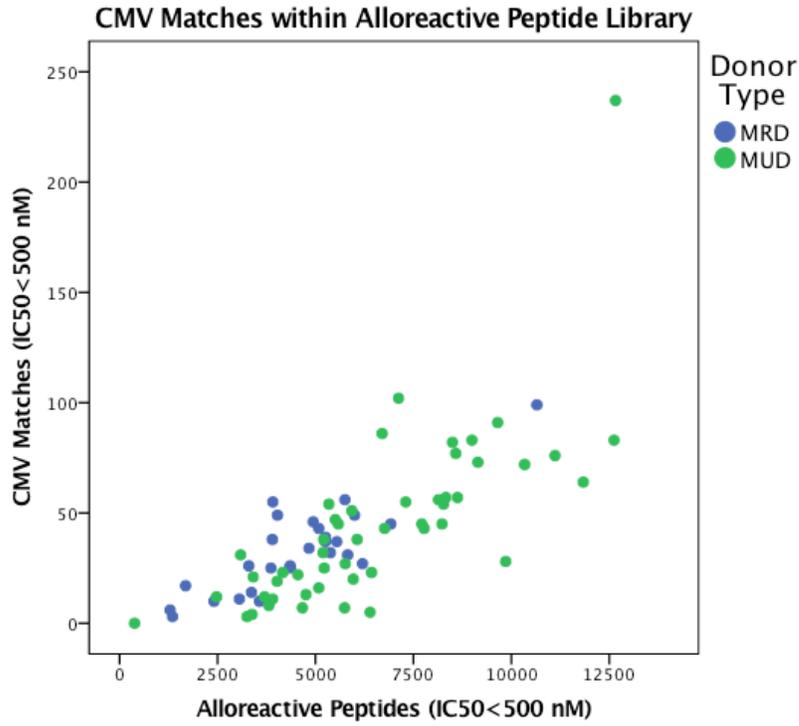

**Figure 4.** Binding affinity similarity indicated by inverse IC50 values from matched human and CMV peptides bound to the same HLA. Each data point represents the intersection of a matched peptide bound to an HLA class I molecule (CMV➔Human). These peptides may be cross reactive, with varying degrees of T cell cross reactivity potential for alloreactivity trigger to ensue (towards the origin on the human peptide axis being the greatest).

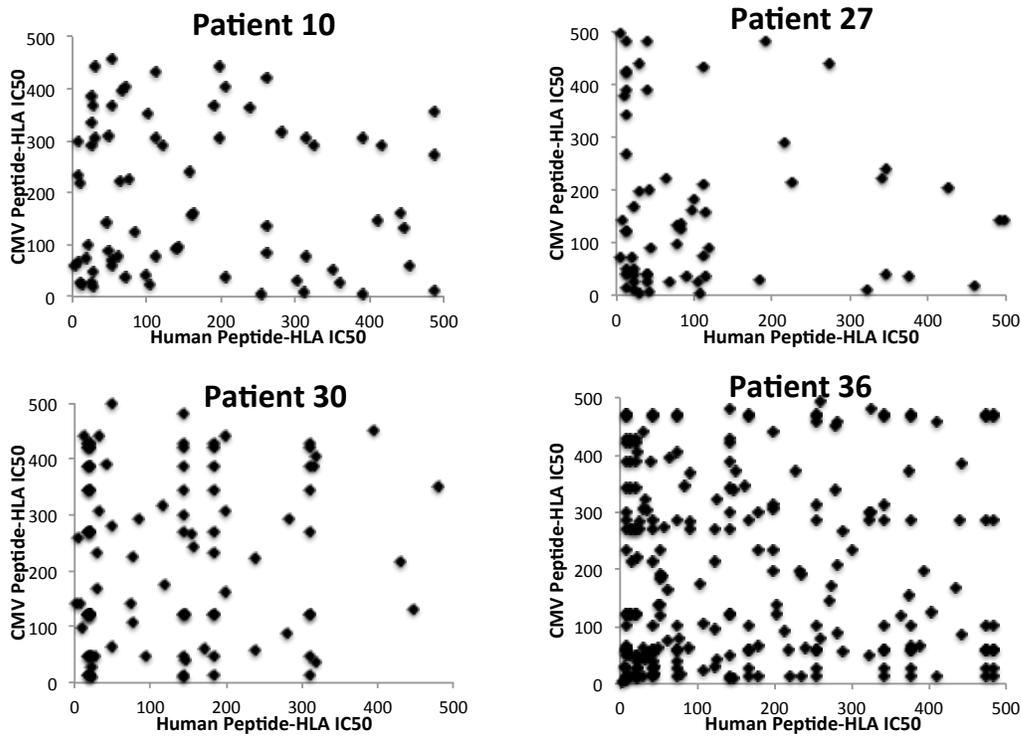

**Figure 5A.** CMV-reactivation course with acute GVHD onset and progression: Patients 47, 68 and 70. Patient 47 exhibited Grade IV GVHD of the GI tract (gut) and the other patients developed steroid refractory Grade IV gut GVHD with other organ involvement. All three patients showed signs of CMV reactivation and bursts of lymphocytosis prior to GVHD onset during stable immunosuppression (TAC, Tacrolimus; SIR, Sirolimus) as measured by serum levels or following taper.

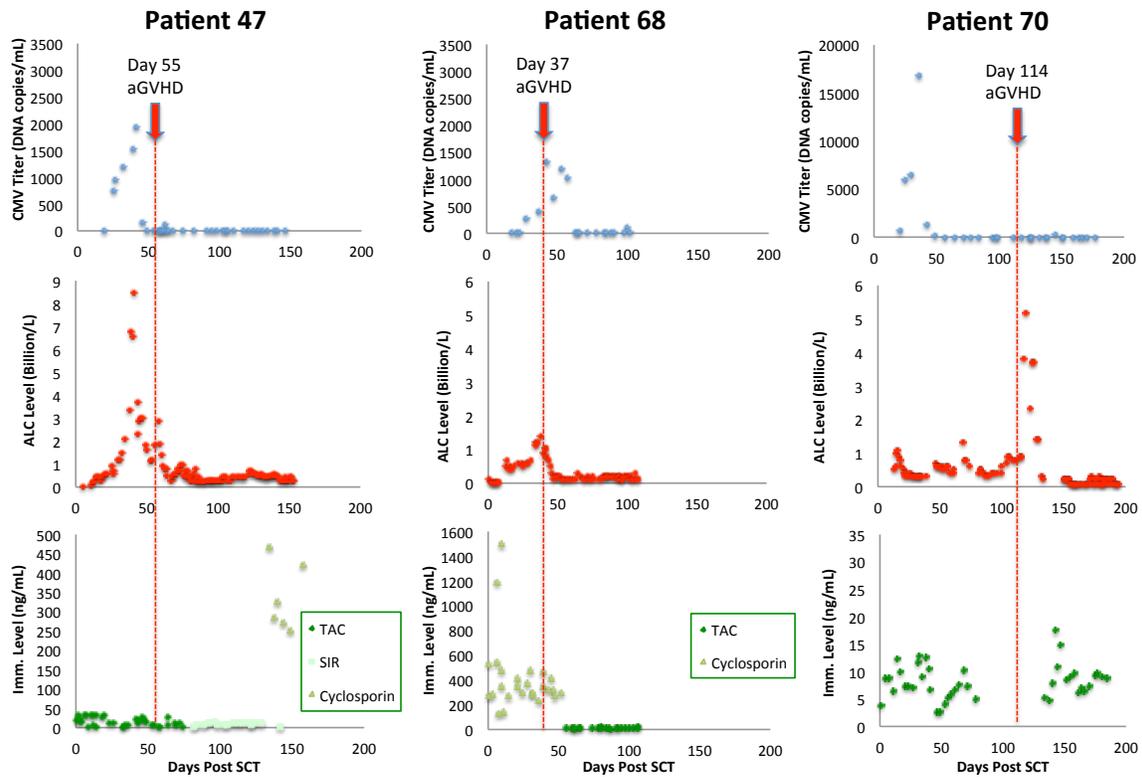

**Figure 5B.** CMV-reactivation with chronic GVHD onset: Patients 10 and 27. Patient 10 had a small spike reactivation prior to a more gradual lymphocyte proliferation during stable immunosuppression levels (TAC, Tacrolimus) and would eventually develop relapsed disease. Patient 27 exhibited a sharp reactivation spike prior to lymphocytosis at stable tacrolimus levels; mainly skin and/or mouth GVHD prior to DLI for both.

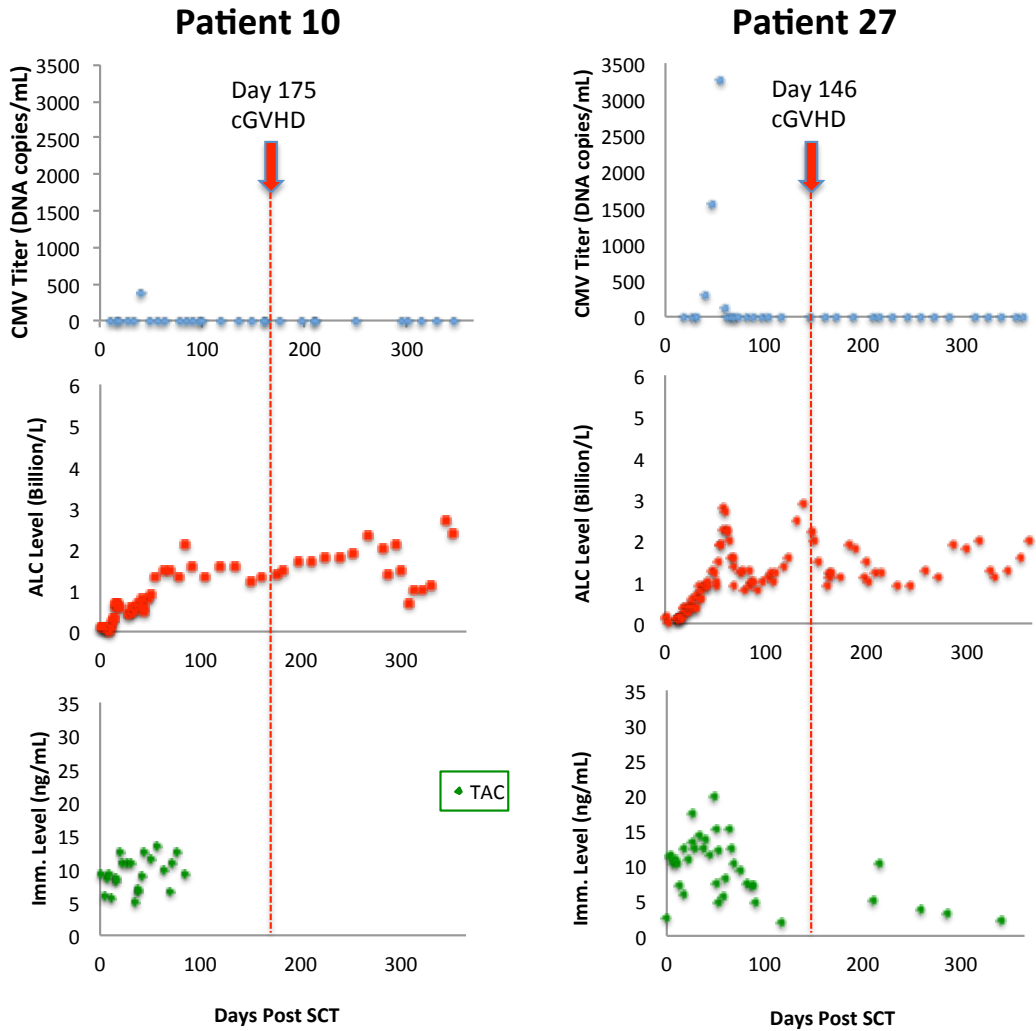

**Figure 5C.** CMV reactivation/de novo infection in patients with aGVHD + cGVHD: Patients 71 (new, continuous infection for 200+ days) and 84 (reactivation). Patient 71 showed poor lymphoid recovery, continuous low CMV infection and eventual relapsed disease. Patient 84 exhibited low-level lymphocyte reactions following CMV reactivation spikes and eventually developed recurrent gut and skin Grade IV GVHD. Both patients received tacrolimus.

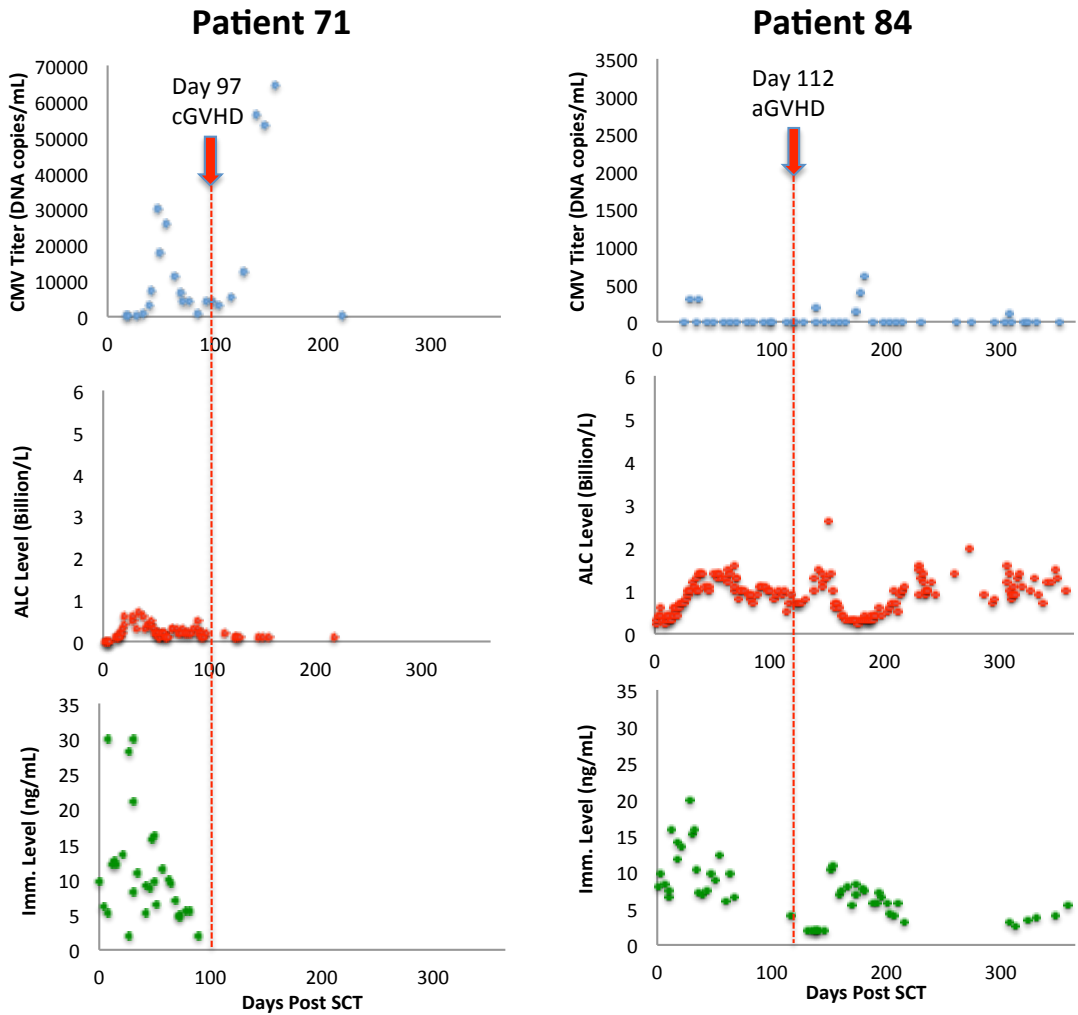

**Supplementary Table 1.** Patient characteristics

| Supplemental Table 1. | Patient Characteristics, n (%) | |
|---|---|---|
| Total Transplants | | 77 (100) |
| Gender | | |
|     Male/Female | | 43 (55.8)/34 (44.2) |
| Patient Age | | |
|     Median (range) | | 55.6 (21-73) |
| | | Total |
| **CMV constellation** | All patients | 77 (100) |
| | D+/R+ | 26 (34) |
| | D-/ R+ | 7 (9) |
| | D+/R- | 13 (17) |
| | D-/ R- | 31 (40) |
| **CMV Prophylaxis** | Acyclovir | 56 (73) |
| | Acyclovir/Valacyclovir | 12 (16) |
| | Valacyclovir | 9 (12) |
| **Age Distribution** | <40 | 12 (16) |
| | 40-59 | 43 (56) |
| | ≥60 | 22 (29) |
| **Donor** | MRD | 26 (34) |
| | MUD | 41 (53) |
| | MMRD | 1 (1) |
| | MMUD | 9 (12) |
| **Stem Cell Source** | Bone Marrow | 7 (9) |
| | Peripheral Blood | 70 (91) |
| **Diagnosis** | Acute lymphoid leukemia | 5 (6) |
| | Acute myeloid leukemia | 29 (38) |
| | Chronic lymphocytic leukemia and lymphomas | 15 (19) |
| | Multiple myeloma | 4 (5) |
| | Myelodysplastic syndromes | 24 (31) |
| **Conditioning Regimen** | Anti-thymocyte globulin (ATG) | 62 (81) |
| | Reduced Intensity | 46 (60) |
| |   ATG/TBI | 19 (25) |
| |   Busulfan/Fludarabine | 27 (35) |
| |   Fludarabine/Melphalan | 2 (77) |
| | Myeloablative | 31 (40) |
| |   Busulfan/Cyclophosphamide | 17 (22) |
| |   Cyclophosphamide/TBI | 11 (14) |
| |   Etoposide/TBI | 1 (1) |
| **GVHD prophylaxis** | Cyclosporin A/Methotrexate | 16 (21) |
| | Cyclosporin A/MMF | 4 (5) |
| | Tacrolimus/Methotrexate | 34 (44) |
| | Tacrolimus/MMF | 23 (30) |
| **GVHD, Overall (n=49)** | Acute, Grades I-II | 20 (26) |
| | Acute, Grades III-IV | 14 (18) |
| | Chronic | 39 (51) |
| | Both, Acute + Chronic | 25 (32) |
| **CMV & GVHD** | CMV, Before GVHD Onset | 13 (17) |
| | CMV, After GVHD Onset | 5 (6) |
| | CMV, No GVHD | 12 (17) |
| | No CMV, GVHD | 31 (40) |
| | No CMV, No GVHD | 16 (20) |

**Supplemental Figure 1**. CMV-seropositive adjusted survival by CMV incidence: GVHD-independent effects of CMV post SCT

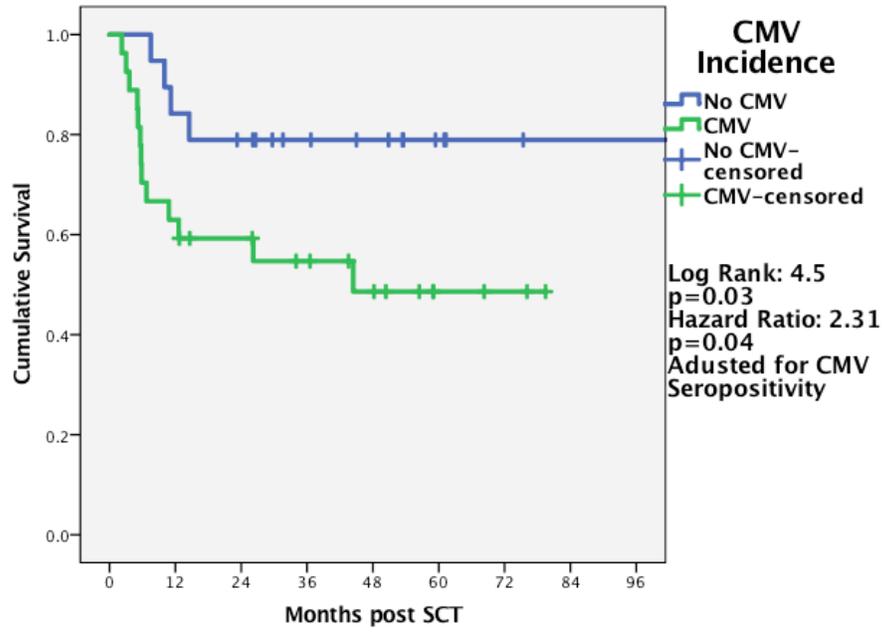

**Appendix List:**

**Appendix A**. Excel workbook – Sheet 1. CMV Proteome – NCBI accession list for hCMV/HHV5

Sheet 2. CMV Immunogenic Protein – CD8+ List from Ref. #27

**Appendix B**. Excel workbook – Sheet 1. Match Analysis Summary – Total peptide # by patient

Sheet 2. Match Analysis Peptides – Screened, sorted pt. peptides

For access to complete appendices or compiled CMV-Human peptide datasets, please contact the first author, Charles Hall, by email at hallce3@vcu.edu for further information.